\documentstyle[aps, preprint]{revtex}
\begin{document}
\preprint{gr-qc/9907031} \draft\tightenlines

\title{Semiclassical Limit and Time in Quantum Cosmology}
\author{Dongsu
Bak\footnote{Electronic address: dsbak@mach.uos.ac.kr}$^{a}$, Sang
Pyo Kim\footnote{Electronic address:
sangkim@knusun1.kunsan.ac.kr}$^{b}$, Sung Ku
Kim\footnote{Electronic address: skkim@theory.ewha.ac.kr}$^{c}$,
Kwang-Sup Soh\footnote{Electronic address:
kssoh@phya.snu.ac.kr}$^{d}$, and Jae Hyung Yee\footnote{Electronic
address: jhyee@phya.yonsei.ac.kr}$^{e}$}
\address{a Department of Physics, University of Seoul, Seoul
130-743 Korea\\ b Department of Physics, Kunsan National
University, Kunsan 573-701 Korea\\ c Department of Physics, Ewha
Women's University, Seoul 120-750 Korea\\ d Department of Physics
Education, Seoul National University, Seoul 151-742 Korea\\ e
Department of Physics, Yonsei University, Seoul 120-749 Korea}

\date{\today}

\maketitle

\begin{abstract}
We propose a method  to recover the  time variable and  the
classical evolution  of the Universe from   the minisuperspace
wave function   of the  Wheeler-DeWitt  equation. Defining a
Hamilton-Jacobi characteristic function $W$ as the imaginary part
of the $\ln \Psi$ we can recover the classical solution, and
quantum corrections. The key idea is to let the energy of the
Wheeler-DeWitt equation vanish only after the semiclassical limit
is taken.
\end{abstract}

In quantum cosmology the universe is  described by a wave
functional rather than by a classical spacetime. The  wave
functional of the universe satisfies  the Wheeler-DeWitt (WDW)
equation $\hat{\cal H} (\pi_{AB}, h_{AB}) \Psi (h_{AB}) = 0$,
where $h_{AB}$ represents three-geometries modulo diffeomorphisms
and $\pi_{AB}$ is its conjugate momentum. A prominent feature of
the WDW equation different from the Schr\"{o}dinger equation is
that a time parameter does not appear to evolve the wave
functionals. It also distinguishes the WDW equation from the field
equations of classical gravity whose evolution is described with
respect to the time variable of spacetime. This is one of the most
conceptually and technically difficult problems in quantum
cosmology \cite{kuchar,isham}.

There have been many proposals on how to define `time' in quantum
cosmology (for review and references, see \cite{kuchar,isham}).
One of these proposals is to select a certain degree of freedom in
the WDW equation as an intrinsic time-like variable. For instance,
the volume factor of the Universe is frequently used and a matter
field is sometimes selected as the intrinsic time-like variable
with respect to which the evolution of the other degrees of
freedom is described.

Another proposal is to extend the superspace by introducing an
additional conjugate pair of an external time and energy. Then the
WDW equation takes the form of the time-dependent
Schr\"{o}dinger-like equation with respect to the external time in
analogy to quantum mechanics. However, an undesirable feature of
this external time approach is that the Schr\"{o}dinger-like
equation leads to a trivial energy eigenvalue equation and the
true time evolution does not appear explicitly.

Still another approach is to define a cosmological (WKB) time as
the directional derivative along the peak of oscillatory wave
functional \cite{kiefer}. The wave functional of the WDW equation
is highly oscillatory in the semiclassical regime, whose peak
follows the Hamilton-Jacobi equation and is thus located along the
classical trajectory. Hence the cosmological time gains its
meaning through a functional relation among the gravitational
degrees of freedom with the back-reaction of matter field taken
into account.

In this letter we propose a new method that unifies the external
time and the Hamilton-Jacobi action. This method enables us to
regain the classical time variable and evolution of the universe
from the wave functional using the correspondence principle which
was well-known in ordinary quantum mechanics. First we present the
essential idea of the method and then apply it to a quantum
Friedmann-Robertson-Walker (FRW) cosmological model.

The WDW equation, through the Dirac quantization procedure, from
the Hamiltonian constraint of classical gravity takes away `time
variable' from the quantum cosmology and makes it difficult to get
the classical cosmology from the wave functional.  We address this
problem not by modifying the Dirac's standard procedure but by
simply observing the final equation, {\it i.e.},
\begin{equation}
\hat{\cal H} \Psi = 0.
\end{equation}
We observe that the set of the solutions $\{\Psi\}$ of the
Hamiltonian constraint is a subset of a larger set which is
defined as the solution space of the equation
\begin{equation}
\label{eigen} \hat{\cal H}\Psi_E = E \Psi_E ,\hspace{10mm}
\forall_E \in R.
\end{equation}
It is a simple mathematical fact that
\begin{equation}
\{ \Psi \} \subset \{\Psi_E \vert E \in R \}.
\end{equation}
Note that the equation (\ref{eigen}) is not a physical one, and is
not according to the procedure of Dirac's quantization, which does
not, however, preclude us from considering the larger set
$\{\Psi_E\}$ as an extended space containing the set of the
solutions of the Hamiltonian constraint.

What we observe is that $\{\Psi_E\}$ of the WDW equation in
minisuperspaces has physical significances, {\it i.e.}, it gives
the correct semi-classical limit of quantum cosmology (`correct'
in the sense that it gives the known classical solution) which is
nothing but the quantum-classical correspondence of the WDW wave
function and the Hamilton-Jacobi characteristic function. Having
obtained the semiclassical limit of the quantum cosmological wave
function it is only natural that the time variable appearing in
the semi-classical formula is taken as a cosmological time.

In order to be concrete we proceed our presentation with simple
examples. To be more specific we confine our study to the closed
FRW universe, a minisuperspace model. The metric for the FRW
universe has only one variable, the scale factor $a(t)$, such as
\begin{equation}
ds^2 = - dt^2 + a^2 d\Omega_3^2
\end{equation}
and a typical evolution for $a$ is given by the Hamiltonian
constraint
\begin{equation}
p^2 + a^2 - \frac{a^4}{a^2_0} = 0.
\end{equation}
The corresponding minisuperspace WDW equation is given by
replacing the momentum $p$ by a differential operator, $p
\rightarrow -i \frac{d}{da}$,
\begin{equation}
\Biggl[\frac{d^2}{da^2} - a^2 +  \frac{a^4}{a_0^2}\Biggr] \Psi (a)
= 0. \label{wd eq}
\end{equation}
Equation (\ref{wd eq}) has been used to propose various quantum
cosmological scenarios \cite{hartle-hawking,linde,vilenkin} and
its interpretation has still aroused much debate in connection
with the boundary conditions \cite{vilenkin2}. Besides there has
been the problem of regaining from the WDW equation and $\Psi(a)$
the properties of the classical universe. In particular, the
'time' variable which is highly conspicuous in the classical
expanding universe does not appear at all in the wave function
$\Psi(a)$.

In order to recover the classical time and evolution from quantum
theory it is worthy to notice that the Hamilton-Jacobi equation of
a particle
\begin{equation}
\frac{1}{2m} \Bigl(\nabla S(q, t) \Bigr)^2 + V (q) +
\frac{\partial S (q, t)}{\partial t} = 0,
\end{equation}
where $S$ is the principal function and $V$ is the potential, is a
short wavelength limit or the semiclassical $(\hbar \rightarrow
0)$  limit of the Schr\"{o}dinger equation
\begin{eqnarray}
\Biggl[ - \frac{\hbar^2}{2m} \nabla^2 + V (q) \Biggr] \Psi (q, t)
= \frac{\hbar}{i} \frac{\partial \Psi (q, t)}{\partial t},
\nonumber\\ \Psi (q, t) = F (q, t) e^{i \frac{S (q, t)}{\hbar}}.
\end{eqnarray}
The principal function can also be obtained from the wave function
by taking the semiclassical limit
\begin{equation}
S (q, t) = \lim_{\hbar \rightarrow 0} \hbar {\rm Im} \Bigl[\ln
\Psi (q, t) \Bigr].
\end{equation}
Especially, the classical trajectory of the particle in a
time-independent potential is obtained from the characteristic
function $W_c (q, E)$ given by
\begin{equation}
S = W_c (q, E) - E t
\end{equation}
as
\begin{equation}
t + \beta = \frac{\partial W_c (q, E)}{\partial E},
\end{equation}
where $E$ is  the energy  of the  particle and $\beta$  is to be
determined by  an initial condition.

In the WDW equation and the wave function $\Psi(a)$ there appears
no energy variable '$E$'  because $E = 0$ in quantum cosmology,
and  hence the procedure to obtain the classical trajectory can
not be directly  applied. Our proposal is that we work with the
non zero-energy wave equation, and let the energy vanish only
after the necessary semiclassical limit is taken. That is, we
begin with the non zero-energy WDW equation
\begin{equation}
\Biggl[\frac{d^2}{da^2} - a^2 + \frac{a^4}{a_0^2} + E \Bigg] \Psi
(a, E) = 0,
\end{equation}
and then take the semiclassical limit of the solution
\begin{equation}
\Psi_0 (a, E) \equiv \Psi(a, E) \Bigl|_{\hbar \rightarrow 0}.
\end{equation}
The characteristic function defined by
\begin{equation}
W_0 (a, E) \equiv \hbar {\rm Im} \Bigl[\ln \Psi_0  (a, E) \Bigr]
\end{equation}
leads to the classical evolution of the universe with respect to
the desired time
\begin{equation}
t + \beta = \frac{\partial}{\partial E} W_0 (a, E) \Bigl|_{E
\rightarrow 0}.
\end{equation}
Now it is natural to consider quantum corrections by letting
\begin{eqnarray}
W (a, E) \equiv \hbar {\rm Im} \Bigl[\ln \Psi (a, E) \Bigr],
\label{w eq}\\ t + \beta = \frac{\partial}{\partial E} W (a, E)
\Bigl|_{E \rightarrow 0}. \label{time}
\end{eqnarray}
It should be commented that the function $W(a, E)$ takes into
account only some part of quantum corrections since the real part
of the wave function is not considered. To include the full
quantum corrections both equations for the imaginary and real
parts of the wave function should be treated. The de Broglie-Bohm
interpretation of wave function is such a method \cite{holland}.
Rather than regarding the equation for the real part as the
conservation law for the probability current, one may solve these
coupled nonlinear equations at the same time to find all the
quantum corrections to the Hamilton-Jacobi equation \cite{kim}.
However, the simple prescription by $W(a, E)$ still gives the
correct classical evolution of the universe and the emergence of
time in the semiclassical limit.

In order to illustrate our arguments we consider a simplified WDW
equation, which is not realistic but shows clearly the points of
this paper. We replace the potential $V (a) = a^2 -
\frac{a^4}{a_0^2}$ by an analytically solvable one
\begin{eqnarray}
V_c (a) = \cases{ 0, & $a < a_1$, \cr V_1~~ (V_1 > 0), & $a_1 < a
< a_2$, \cr - V_2~~ (V_2 > 0), & $a_2 < a$. \cr}
\end{eqnarray}
The wave function under the potential barrier is given by
\begin{eqnarray}
\Psi = \cases{A e^{i k_I a} + B e^{-ik_I a}, & $a < a_1$, \cr
Ce^{- \kappa_{II} a} + D e^{\kappa_{II} a}, & $a_1< a < a_2$, \cr
F e^{i k_{III} a} + G e^{-ik_{III} a}, & $a_2 < a$, \cr}
\end{eqnarray}
where $A, B, C, D, F, G$ are appropriate constants, and
\begin{equation}
k_I = \frac{\sqrt{2E}}{\hbar},~~\kappa_{II} = \frac{\sqrt{2(V_1 -
E)}}{\hbar},~~k_{III} = \frac{\sqrt{2(E + V_2)}}{\hbar}.
\end{equation}
In the case of only outgoing wave function $(G = 0)$ in the
out-region $a > a_2$, we have
\begin{equation}
W (a, E) = \hbar {\rm Im} \Bigl[ \ln \Psi (a, E) \Bigr] = a
\sqrt{2 (E + V_2)} + {\rm constant},
\end{equation}
which is in fact identical to the classical Hamilton-Jacobi
characteristic function $W_c$. Hence in the simplified model the
quantum wave function and the classical characteristic function
give the identical evolution of the universe
\begin{equation}
t + \beta = \frac{\partial W}{\partial E} \Bigl|_{E = 0} =
\frac{a}{\sqrt{2V_2}}.
\end{equation}
Thus the outgoing wave function gives a linearly expanding
universe, and similarly the incoming wave gives a linearly
collapsing universe.

Regaining the time variable and the classical evolution of the
universe using our method does not work in the tunneling region
where ${\rm Im} \Bigl[\ln \Psi (a, E) \Bigr]$ does not have a
classical counterpart. It is not a failure of our prescription,
though, because in the tunneling situation there does not exist a
classical solution, and hence no evolution and no relevant time
variable. What is relevant is the tunneling probability which can
be read from the wave function directly, and hence the classical
time variable is irrelevant and unnecessary. Nevertheless, we also
note that the WKB approximation $P = e^{-2 [W_c (a_2) - W_c
(a_1)]} = e^{- 2 \kappa_{II} (a_2 - a_1)}$ can be obtained as a
classical limit from $P = e^{2 \hbar {\rm Re} [\ln \Psi(a_2) - \ln
\Psi (a_1) ]}$, which implies that in this region $W_0 (a, E) =
\hbar {\rm Re} \Bigl[ \ln \Psi (a, E) \Bigr] $ rather than the
imaginary part.

The Hartle-Hawking's no-boundary \cite{hartle-hawking} and the
Linde's wave functions \cite{linde} of the universe are linear
superpositions of expanding and collapsing branches, whereas the
Vilenkin's wave function \cite{vilenkin} describes one branch of
expanding universe. In the former wave functions a direct
application of the function ${\rm Im} \Bigl[ \ln \Psi(a) \Bigr]$
will involve interference terms, and the simple identification
with the characteristic function $W$ may not be straightforward.
However, for each branch of wave function, either expanding or
collapsing universe, we recover the semiclassical limit. When one
component (say, expanding one) is dominant we may still apply our
method to recover the classical evolution of the Universe,
treating the interference terms as quantum corrections. This view
point can also be used in quantum treatments of black hole
formation by collapse of a massless scalar field \cite{bak}.

\begin{acknowledgements}
We would like to appreciate the warm hospitality of CTP of Seoul
National University where this paper was completed. This work was
supported in parts by BSRI Program under BSRI 98-015-D00054,
98-015-D00061, 98-015-D00129. DB was supported in part by KOSEF
Interdisciplinary Research Grant 98-07-02-07-01-5, SPK by the
International Program of Korea Research Foundation, 1997 and JHY
by KOSEF under Grant No. 98-07-02-02-01-3.
\end{acknowledgements}

\end{document}